\documentclass[12pt]{iopart}

\usepackage{graphicx}
\usepackage{amssymb}

\usepackage{graphicx}
\usepackage{graphicx}
\usepackage{color}
\usepackage{pstricks, pst-plot, pst-all}
\usepackage{units}

\begin{document}

\title{Physics on Smallest Scales - An Introduction to Minimal Length Phenomenology}
\author{Martin Sprenger$^{1,2}$, Piero Nicolini$^1$ and Marcus Bleicher$^1$}
\address{$^1$Institute for Theoretical Physics, Goethe University and Frankfurt Institute for Advanced Studies, Ruth-Moufang-Str.\ 1, 60438 Frankfurt am Main, Germany\\
$^2$DESY Theory Group, DESY Hamburg, Notkestr.\ 85, 22603 Hamburg, Germany}
\ead{\mailto{sprenger@fias.uni-frankfurt.de}\\\mailto{nicolini@th.physik.uni-frankfurt.de}\\\mailto{bleicher@th.physik.uni-frankfurt.de}}

\begin{abstract}

Many modern theories which try to unify gravity with the Standard Model of particle physics, as e.g. string theory, propose two key modifications to the commonly known physical theories:
\begin{itemize}
\item \ the existence of additional space dimensions;
\item \ the existence of a minimal length distance or maximal resolution.
\end{itemize}
While extra dimensions have received a wide coverage in publications over the last ten years (especially due to the prediction of micro black hole production at the LHC), the phenomenology of models with a minimal length is still less investigated.
In a summer study project for bachelor students in 2010 we have explored some phenomenological implications of the potential existence of a minimal length.
In this paper we review the idea and formalism of a quantum gravity induced minimal length in the generalised uncertainty principle framework as well as in the coherent state approach to non-commutative geometry.
These approaches are effective models which can make model-independent predictions for experiments and are ideally suited for phenomenological studies.
Pedagogical examples are provided to grasp the effects of a quantum gravity induced minimal length.
This article is intended for graduate students and non-specialists interested in quantum gravity.
\end{abstract}

\vspace{1.5cm}
\small{DESY 12-014}

\maketitle

\section{Introduction}
Quantising gravity is one of the most important problems in modern theoretical physics.
Even though people have been working on it for more than 50 years, there is no complete theory of quantum gravity up to this date.
A major obstacle in finding the correct theory is the absence of any experimental quantum gravity signal.
Out of this frustration, a new research field has emerged in the last decade which employs effective theories to describe quantum gravity effects and to look for possible experimental signatures.
This field is called quantum gravity phenomenology.
The effective theories are built by using standard quantum field theory or general relativity and implementing one or several features that are supposed to arise in a full theory of quantum gravity.
As there are effects which appear in the majority of candidate theories, quantum gravity phenomenology can implement these effects and make predictions independently of any fundamental formulation.
Furthermore, predictions can be made without employing the heavy machinery of string theory or loop quantum gravity from which it is usually very difficult to foresee experimental signatures.

For these reasons we believe that quantum gravity phenomenology is suitable to be presented to a large audience. The purpose of this paper is to provide an accessible explanation of the basic ideas and results (at least a part of them) of quantum gravity phenomenology in order to keep the scientific community informed about the progress in a field that too often is out of the understanding of non-specialists. More importantly, we aim to address our work not only to non-specialists but also to students with a background at the level of introductory courses in quantum mechanics, special relativity and particle physics (in Germany this is often at the reach of undergraduate students, but we safely restrict the readership to the graduate ones). 

The effect we want to study in this article is the appearance of a minimal length.
The fact there is a radical change in the nature of space-time comes as no surprise.
In fact, we have to take into account that quantum field theory and general relativity treat space-time very differently.
While general relativity treats space-time as a dynamical entity using the metric tensor $g_{\mu\nu}$, in quantum field theory space and time are mere labels on which fields are defined.\\
The idea of a minimal length dates far back, long before the birth of modern candidate theories of quantum gravity (cf.\ \cite{Snyder1947}) and appears in all recent formulations like the path-integral quantisation \cite{Padmanabhan1985}, string theory \cite{Gross1988}, loop quantum gravity \cite{Rovelli1995} and other approaches \cite{Mead1964}, \cite{Doplicher1995}.
To motivate the emergence of a minimal length, recall from optics that to probe a structure of length $\lambda$ one needs photons of wavelength $\lambda$ or less.
To resolve smaller and smaller structures, the energy of the photons (which is proportional to the inverse of the wavelength, $\lambda\sim\frac{1}{E_{\gamma}}$) has to be increased.
However, when gravity is taken into account, there is the possibility that the energy density is large enough to create a black hole and all information on the structure one wants to probe is lost behind the event horizon.
Therefore, a fundamental (minimal) length scale naturally emerges in any quantum theory in the presence of gravitational effects that accounts for a limited resolution of space-time.
As there is only one natural length scale we can obtain by combining gravity ($G$), quantum mechanics ($\hbar$) and special relativity ($c$), this minimal length is expected to appear at the Planck scale, 
\begin{equation}
\ell_P\equiv\sqrt{\frac{\hbar G}{c^3}}\approx 10^{-35}~\mathrm{m}.
\end{equation}
If the minimal length were of that order, current and near-future experiments would have rather remote chances to observe quantum gravity effects.
However, our description of gravity has only been tested down to length scales of the order of $0.1~\mathrm{mm}$ by direct measurements of Newton's law \cite{Hoyle2001} and down to $1~\left(\hbar/\mathrm{TeV}\right)\approx 10^{-19}~\mathrm{m}=10^{-4}~\mathrm{fm}$\footnote{In this paper we adopt the units $c=1$, while keeping $\hbar$ in the formulas to emphasize the quantum nature of the formalism.
For readers not familiar with this system we briefly recall that for $c=1$ velocities turn into dimensionless quantities, lengths and times have the same dimension, and masses and momenta have the same dimension as energies.
As a final note we recall that the acronym TeV stands for teraelectronvolt, corresponding to $10^{12} \mathrm{eV}\approx 1.602 \times 10^{-7}\mathrm{J}$, the energy scale currently under scrutiny of the Large Hadron Collider (LHC) at CERN laboratories, Geneva. Macroscopically, $1~\mathrm{TeV}$ corresponds to a temperature $T\approx 1.602\times 10^{-7}\mathrm{J}k_B^{-1}\approx 1.16 \times 10^{16} K$.
} by indirect searches at particle colliders \cite{Nakamura2010}.
Therefore the minimal length might lie anywhere between the Planck scale and $10^{-4}~\mathrm{fm}$.
In the rest of the article we want to introduce an effective theory modeling a minimal length and study two scenarios in this framework.

\section{Generalised Uncertainty Principle}
Let us now discuss how to introduce a minimal length into standard quantum mechanics by the so-called generalised uncertainty principle (GUP).
Pioneering work on this approach can be found in \cite{Kempf1995}, \cite{Kempf1997}.
One of the fundamental concepts in quantum mechanics is the Heisenberg uncertainty principle (see, for example, \cite{Sakurai1993})
\begin{equation}\Delta x\Delta p\geq \hbar,\label{eq:heisenberg}\end{equation}
where $\Delta x$ is the uncertainty in position and $\Delta p$ is the uncertainty in momentum.
The Heisenberg uncertainty principle prevents us from measuring the position and the momentum of a particle simultaneously to arbitrary precision.
However, it is always possible to measure the position of a particle with better and better accuracy if one compensates for that by making the momentum uncertainty larger and larger.
As an extremal case, particles in momentum eigenstates (i.e.\ with vanishing momentum uncertainty) are completely delocalised.
Now, imagine Eq.(\ref{eq:heisenberg}) had an extra term proportional to $(\Delta p)^2$ on the right-hand side
\begin{equation}\Delta x\Delta p\geq \hbar(1+\beta (\Delta p)^2).\label{eq:mod_heisenberg}\end{equation}
The above argument for position measurements with increasing accuracy runs into problems, because increasing $\Delta p$ makes the right-hand side grow even faster than the left-hand side.
At one point, it will not be possible to fulfill the inequality anymore, therefore there is a minimal $\Delta x \neq 0$ for which the inequality holds.
This means there is a maximal resolution of position or, in turn, a minimal length.
The introduction of the extra term looks artificial, but this exact form of the modification arises for example in string theory \cite{Konishi1990}.
In quantum mechanics, such uncertainty relations usually follow from the non-commutativity of the corresponding operators.
For example, Eq.(\ref{eq:heisenberg}) follows from
\begin{equation}\left[\hat{x},\hat{p}\right]=i\hbar,\end{equation}
where $[\cdot,\cdot]$ is the commutator, and in general
\begin{equation}\Delta A\Delta B\geq \frac{1}{2}\left\langle\left[\hat{A},\hat{B}\right]\right\rangle,\end{equation}
where $\langle...\rangle$ denotes the expectation value.
The modified Heisenberg uncertainty relation Eq.(\ref{eq:mod_heisenberg}) therefore follows from a commutator
\begin{equation}\left[\hat{x},\hat{p}\right]=i\hbar(1+\beta \hat{p}^2)\label{eq:commutator_gup}\end{equation}
and can be obtained by an extension of the usual operator representation in momentum space, $\hat{x}=i\hbar\partial_p$, to
\begin{equation}\hat{x}=i\hbar (1+\beta p^2)\partial_p.\end{equation}
The momentum operators remain unchanged for simplicity\footnote{One can also investigate theories with non-commuting momenta, see e.g. \cite{Hinrichsen1996},\cite{Kober2010}.}.
This representation already includes the effects of a minimal length.
It is therefore possible to recalculate standard quantum mechanical problems and find modifications due to the minimal length.\\
The commutator Eq.(\ref{eq:commutator_gup}) belongs to a larger class of commutators of the form
\begin{equation}\left[\hat{x}_i,\hat{p}_j\right]=i\hbar\delta_{ij}(1+f(p^2)),\label{eq:gen_comm}\end{equation}
where $f(p^2)$ is a generic function which has to vanish for small momenta and should preserve symmetries such as rotations but is otherwise arbitrary.
The most general momentum space representation of $\hat{x}_i$ then looks like
\begin{equation}\hat{x}_i=i\hbar (1+f(p^2))\partial_{p_i}.\end{equation}
Recall from quantum mechanics that
\begin{equation}\left[\left[\hat{x}_i,\hat{x}_j\right],\hat{p}_k\right]+\left[\left[\hat{x}_j,\hat{p}_k\right],\hat{x}_i\right]+\left[\left[\hat{p}_k,\hat{x}_i\right],\hat{x}_j\right]=0.\label{eq:jacobi}\end{equation}
This is not a mere coincidence but holds for all elements of a mathematical structure called Lie algebra and is called the Jacobi identity.
The position and momentum operators in the GUP model still form a Lie algebra.
Therefore, the position commutator is already fixed by Eqs.(\ref{eq:gen_comm}),(\ref{eq:jacobi}) and reads
\begin{equation}\left[\hat{x}_i,\hat{x}_j\right]=-2i\hbar \left(\hat{x}_i\hat{p}_j-\hat{x}_j\hat{p}_i\right)f'(p^2),\end{equation}
where $f'(p^2)$ is the derivative of $f(p^2)$ with respect to $p^2$.
This commutator looks complicated, but the important point is that it is non-vanishing as long as $f'(p^2)\neq 0$, again indicating a minimal length.\\
Several points should be noted here.
First of all, as shown in \cite{Kempf1995}, by introducing the new commutation relations, the position operator is no longer Hermitian.
Still, one can keep the symmetry of the position operator by introducing a modified momentum integration measure
\begin{equation}d^3p\rightarrow \frac{d^3p}{1+f(p^2)},\label{eq:measure}\end{equation}
because, for states $\left|\psi\right\rangle$, $\left|\phi\right\rangle$ whose wavefunction vanishes at infinity,
\begin{eqnarray}
\nonumber\left(\left\langle\psi\right|\hat{x}\right)\left|\phi\right\rangle&=&\int\limits_{-\infty}^{\infty}\frac{dp}{1+f(p^2)}\psi^*(p)i\hbar(1+f(p^2))\partial_p\phi(p)\\
&=&\int\limits_{-\infty}^{\infty}\frac{dp}{1+f(p^2)}\left(i\hbar(1+f(p^2))\partial_p\psi(p)\right)^*\phi(p)=\left\langle\psi\right|\left(\hat{x}\left|\phi\right\rangle\right)
\end{eqnarray}
as one can check by partial integration.
Note that this operator would not be symmetric if we had not included the compensating factor in the integration measure.
The symmetry of $\hat{p}$ is obvious.
As momenta are still commuting, we can use the momentum eigenbasis as in ordinary quantum mechanics, however, working with the modified integration measure.
A second point concerns the position eigenbasis.
As the theory contains an inherent uncertainty in position, position eigenstates (i.e. states which have zero uncertainty in position) cannot exist in this framework.
The next best thing to use are states with minimal position uncertainty.
In \cite{Kempf1995}, it is shown how to construct these states and the corresponding maximum localisation (quasi-)basis.
For our purposes, however, it is enough to look at momentum eigenstates which still are plane waves, but with a modified dispersion relation
\begin{equation}\lambda(p)=2\pi\hbar\left(\int\frac{dp}{1+f(p^2)}\right)^{-1},\end{equation}
where $\lambda$ cannot be smaller than the minimal length.
Now we have all tools we need to calculate problems at hand, for more details on the GUP model we refer to \cite{Kempf1995} and \cite{Kempf1997}.
\section{Non-commutative Geometry}
Before continuing to study examples, let us briefly mention a different class of effective theories giving rise to a minimal length, non-commutative geometry (NCG).
Instead of starting with a modification in the commutator between position and momentum operators, NCG modifies the commutator between position operators:
\begin{equation}\left[\hat{x}_\mu, \hat{x}_\nu\right]=i\theta_{\mu\nu}, \end{equation}
where $\theta_{\mu\nu}$ is an antisymmetric matrix with entries of dimension of an area that govern the non-commutative behaviour.
There are several possibilities to implement such a theory, the most popular being based on the Moyal product.
In this framework the usual product is replaced by the $\star$-product which can be represented in $4$ dimensions by 
\begin{equation}(f\star g)(x)=\int \frac{d^4yd^4k}{(2\pi)^4} f\left(x^\mu+\frac{1}{2}\theta^{\mu\nu}k_{\nu}\right)g(x^\mu+y^\mu)e^{ik_\nu y^\nu},\end{equation}
where $k^\mu = \left(\omega, \vec{k} \right) \,$ is the wave four-vector (cf. \cite{Wulkenhaar2006}).
This suffers from some technical difficulties which we do not want to describe here, for reviews see \cite{Douglas2001}, \cite{Wulkenhaar2006}.\\
In recent years, another approach to NCG has been developed, the coherent state approach.
In this framework, position eigenstates are replaced by coherent position states.
This is achieved by the action of an operator $e^{\theta\bigtriangleup}$ on the classical position eigenstates $\delta^{(3)}(x)$, where $\theta$ is the minimal length squared and $\bigtriangleup$ is the Laplace operator.
Using the relation
\begin{equation}
e^{\theta\bigtriangleup}\delta^{(3)}(x)=\frac{e^{\theta\bigtriangleup}}{(2\pi)^3}\int d^3p\, e^{\frac{i}{\hbar}\vec{x}\vec{p}}=\frac{1}{(2\pi)^3}\int d^3p\, e^{-\theta p^2/\hbar^2}e^{\frac{i}{\hbar}\vec{x}\vec{p}}
\end{equation}
it is shown in \cite{Smailagic2003a}, \cite{Smailagic2003b} that all modifications can be accounted for by a modified integration measure
\begin{equation}d^3p\rightarrow d^3p~e^{-\theta p^2/\hbar^2}.\end{equation}
This, however, is nothing but the GUP model with 
\begin{equation}f(p^2)=e^{-\theta p^2/\hbar^2}-1,\label{eq:f_ncg}\end{equation}
cf. Eq.(\ref{eq:measure}).
Therefore, these models are equivalent and in the following we will use the GUP model with the choice Eq.(\ref{eq:f_ncg}) for $f(p^2)$.
Plane waves will then have a dispersion relation
\begin{equation}\lambda(p)=\frac{2\pi\sqrt{\theta}}{\frac{\sqrt{\pi}}{2}~\mathrm{erf}\left(\frac{\sqrt{\theta}}{\hbar} p\right)},\label{eq:mod_plane_wave}\end{equation}
where $\mathrm{erf}(x)$ is the error function $\mathrm{erf}(x)=\frac{2}{\sqrt{\pi}}\int\limits_0^x e^{-t^2}dt$.

\section{A simple example...}
\begin{figure}
\centering
\includegraphics{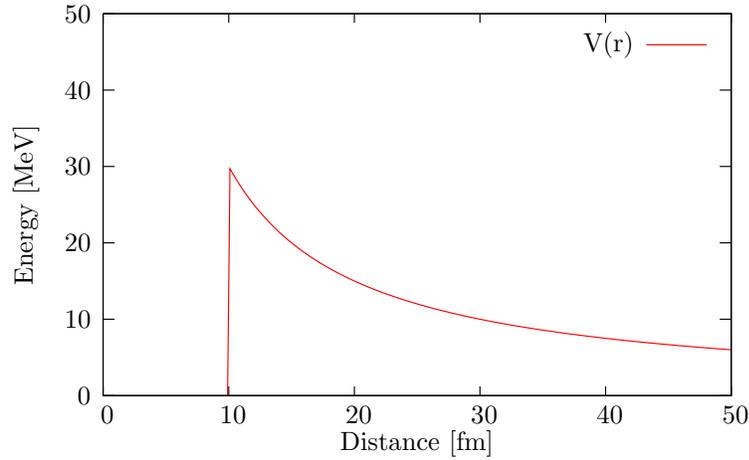}
\caption{Coulomb potential for $\alpha$ decay}
\label{fig:potential}
\end{figure}

\begin{figure}
\centering
\includegraphics{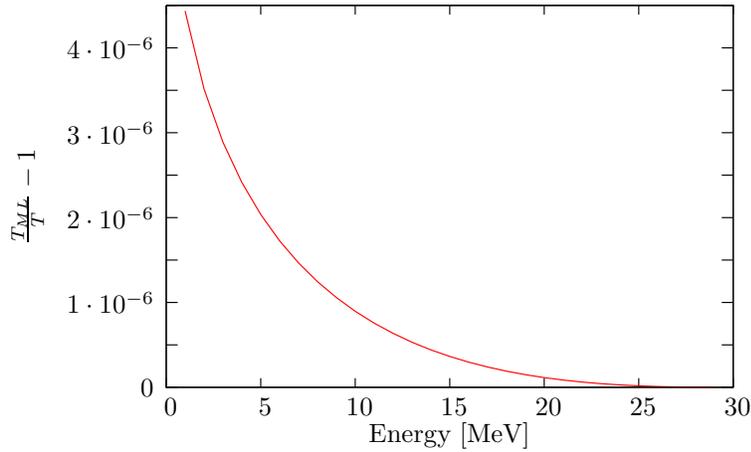}
\caption{Relative difference in the tunneling probability with and without a minimal length, for $V_0=30~\mathrm{MeV}$, $a=10~\mathrm{fm}$ and $\hbar^2/\theta=1~\mathrm{TeV}^{2}$.}
\label{fig:tunnel_prob}
\end{figure}

One of the first problems in quantum mechanics every student has to solve is the potential barrier.
One finds that particles can tunnel through a barrier even though their energy is smaller than the height of the potential.
Applied to a potential of the form
\begin{equation}V(r)=\frac{e^2}{4\pi\epsilon_0}\frac{Z-2}{r}\theta(r-r_0),\end{equation}
with $e$ being the electron charge, $\epsilon_0$ the electric constant, $Z$ the proton number of the nucleus and $r_0$ the radius of the nucleus (see figure \ref{fig:potential}), this potential is used to describe the alpha decay of nuclei.
It already takes into account that two protons are taken from the nucleus by the alpha particle and that the potential is only effective when the alpha particle is outside the nucleus with radius $r_0$.
Starting from the potential barrier, one can approximate the tunneling probability for the Coulomb potential by splitting it up into several rectangular potential barriers and multiplying the probabilities to tunnel through each segment.
For a constant potential $V$, the wavefunction is a momentum eigenstate given by
\begin{equation}
\psi(x)= Ae^{ikx}+Be^{-ikx},
\end{equation}
with $A$,$B$ constants and where $k$ is given by
\begin{equation}
k=\frac{p}{\hbar}=\sqrt{\frac{2m}{\hbar^2}(E-V)}.
\end{equation}
In the limit of infinitely many segments, one obtains the well-known WKB result (see \cite{Sakurai1993})
\begin{equation} T=\frac{\exp\left(-2\int\limits_a^b dx\sqrt{\frac{2m}{\hbar^2}(V(x)-E)}\right)}{\left(1+\frac{1}{4}\exp\left(-2\int\limits_a^b dx\sqrt{\frac{2m}{\hbar^2}(V(x)-E)}\right)\right)^2},\label{eq:transmission}\end{equation}
where $a$ and $b$ are the classical turning points $E(a)=V(a)$, $E(b)=V(b)$ and $T$ is the tunneling probability.
To rederive this result including a minimal length, recall that in the GUP model the momentum basis remains unchanged, except for a modification of the integration measure.
Since the free Hamiltonian is diagonal in momentum space, the solutions in all regions are still momentum eigenstates.
However, as shown before, these are plane waves with a modified dispersion relation.
For a constant potential $V$ this would lead to solutions of the form 
\begin{equation}
\psi(x)= Ae^{ikx}+Be^{-ikx},
\end{equation}
where $k$ is now given by
\begin{equation}
k=\frac{\sqrt{\pi}}{2\sqrt{\theta}}\mathrm{erf}\left(\frac{\sqrt{\theta}}{\hbar}p\right),
\end{equation}
 see Eq.(\ref{eq:mod_plane_wave}).
Therefore, the only modification introduced by the minimal length is the modification of the dispersion relation in Eq.(\ref{eq:transmission}) and the modified transmission probability is given by
\begin{equation} T_{ML}=\frac{\exp\left(-\sqrt{\frac{\pi}{\theta}}\int\limits_a^b dx~\mathrm{erf}\left(\sqrt{\frac{2m\theta}{\hbar^2}(V(x)-E)}\right)\right)}{\left(1+\frac{1}{4}\exp\left(-\sqrt{\frac{\pi}{\theta}}\int\limits_a^b dx~\mathrm{erf}\left(\sqrt{\frac{2m\theta}{\hbar^2}(V(x)-E)}\right)\right)\right)^2}.\label{eq:mod_trans}\end{equation}
Realistic values for the radius of the nucleus and the strength of the potential are $r_0\sim 10~\mathrm{fm}$, $V_0\sim 30~\mathrm{MeV}$.
Numerical integration shows that for the most optimistic case of $\theta/\hbar^2=1~\mathrm{TeV}^{-2}$ the relative difference of the two transmission probabilities is of the order of $10^{-6}$ as shown in figure \ref{fig:tunnel_prob}. The relative difference between the half-lives with and without minimal length is, of course, of the same order of magnitude.
Comparing this to experimental data found in \cite{Audi2003}, we see that this is too small to be detectable with current statistics, which lies between $10^{-1}$ to $10^{-2}$ relative difference.
However, using a large enough system of decaying particles, it does not seem too far-fetched to hope for a bound on the minimal length from this effect in the future.
\\
\section{...and a more complicated one}
\begin{figure}
\centering
\includegraphics{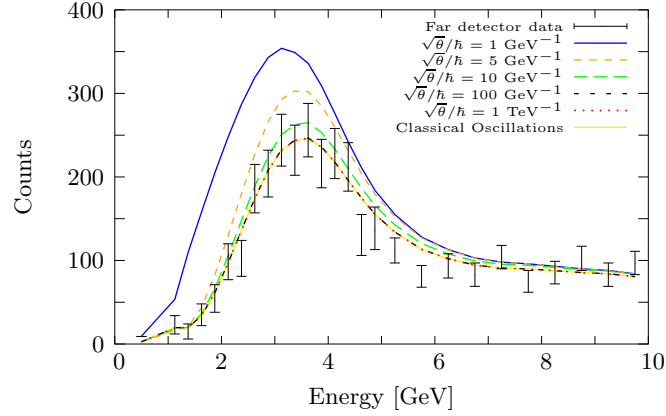}
\caption{Oscillation pattern for classical oscillations and modified oscillations for several values of $\sqrt{\theta}$.}
\label{fig:minos}
\end{figure}

Our best description of particle physics, the Standard Model, consists of three generations and in each generation there are two quarks, a charged lepton and a neutrino.
The neutrinos are named after the corresponding charged lepton, so there are electron neutrinos $\nu_e$, muon neutrinos $\nu_\mu$ and tau neutrinos $\nu_\tau$.
In nuclear processes, neutrinos are created in one of these so-called flavours.
It seems logical that neutrinos remain in the flavour state in which they were created if they are not interacting.
However, neutrinos behave differently.
This was first noted by the Homestake experiment in the 1960s which measured the flux of solar neutrinos (for a review, see \cite{Cleveland1998}).
The measured flux turned out to be too small by a factor of three compared to the theoretical predictions.
It was not until 2001 that the SNO experiment, a neutrino observatory located 
in Sudbury, Canada,  could explain this deficit \cite{SNO2001}.
The Homestake experiment was only sensitive to electron neutrinos and while solar neutrinos are always created as electron neutrinos, SNO registered muon neutrinos and tau neutrinos, as well, so that the total flux met the theoretical predictions.
Therefore, neutrinos seem to change their flavour upon free propagation.
But if neutrinos do not propagate in flavour eigenstates, in eigenstates of which quantity do they propagate?
While the Standard Model treats neutrinos as massless particles, it became clear that the idea of neutrinos having a mass (a tiny mass, that is, current bounds can be found in \cite{Nakamura2010}) could account very well for the flavour oscillations as we will see.\\
For simplicity, we will assume there are only two flavours.
In the following, Latin subscripts stand for mass eigenstates while Greek indices stand for flavour eigenstates.
Then, flavour eigenstates in the flavour eigenbasis are given by
\begin{equation}\left|\nu_\alpha\right\rangle=\left(\begin{array}{c}1\\0\end{array}\right),\quad \left|\nu_\beta\right\rangle=\left(\begin{array}{c}0\\1\end{array}\right).\label{eq:states}\end{equation}
The basis change from the flavour eigenbasis to the mass eigenbasis will be described as usually in quantum mechanics by a unitary matrix $U$, which in two dimensions can be parametrised by a single angle:
\begin{equation}U=\left(\begin{array}{cc}\cos\theta &\sin\theta\\-\sin\theta &\cos\theta\end{array}\right)\label{eq:2fmixing}\end{equation}
As we assume that the free Hamiltonian is diagonal in the mass eigenbasis, we can represent it by the matrix
\begin{equation}\hat{H}=\left(\begin{array}{cc}\sqrt{p^2+m^2_1}&0\\0&\sqrt{p^2+m^2_2}\end{array}\right),\label{eq:h_mass}\end{equation}
where $m_1$ and $m_2$ are the masses of the mass eigenstates.
A neutrino in a mass eigenstate propagating freely can then be represented by a plane wave:
\begin{equation}\left|\nu_k(t)\right\rangle=e^{-\frac{i}{\hbar}E_kt}\left|\nu(0)\right\rangle\label{eq:plane_wave}\end{equation}
Collecting results, we can immediately write down the transition amplitude for a neutrino created in flavour eigenstate $\left|\nu_\alpha\right\rangle$ to be found in flavour eigenstate $\left|\nu_\beta\right\rangle$ after propagating freely for a time t:
\begin{equation}\left\langle\nu_\beta|\nu_\alpha(t)\right\rangle=\left\langle\nu_\beta\right|U^\dagger e^{-\frac{i}{\hbar}\hat{H}t}U\left|\nu_\alpha\right\rangle\end{equation}
Putting in the explicit form of the matrices and the states Eqs.(\ref{eq:states}),(\ref{eq:2fmixing}),(\ref{eq:h_mass}) we find
\begin{equation}
\left\langle\nu_\beta|\nu_\alpha(t)\right\rangle=\frac{1}{2}\sin(2\theta)\left(e^{-\frac{i}{\hbar}E_1t}-e^{-\frac{i}{\hbar}E_2t}\right).
\end{equation}
As the energies are assumed to be much larger than the neutrino masses, the difference of energies $E_1-E_2$ can be approximated by
\begin{eqnarray}
E_1-E_2&=&\sqrt{p^2+m_1^2}-\sqrt{p^2+m_2^2}=|p|\sqrt{1+\frac{m_1^2}{p^2}}-|p|\sqrt{1+\frac{m_2^2}{p^2}}\\
&\approx& \frac{m_1^2-m_2^2}{2E}=\frac{\Delta m^2}{2E},
\end{eqnarray}
where $E=|p|$ and $\Delta m^2=m_1^2-m_2^2$.
From that, we find the transition probability simply as the square of the transition amplitude:
\begin{equation}P(\nu_\alpha\rightarrow\nu_\beta)=\left|\left\langle\nu_\beta|\nu_\alpha(t)\right\rangle\right|^2=\sin^2(2\theta)\sin^2\left(\frac{\Delta m^2}{4\hbar E}t\right).\end{equation}
Indeed, we find an oscillatory behaviour for the neutrino.\\
To include the effect of the minimal length, note that in (\ref{eq:plane_wave}) use was made of the dispersion $\omega_k=\frac{E_k}{\hbar}$.
In the GUP model, this is modified as in Eq.(\ref{eq:mod_plane_wave})
\begin{equation}\omega(E)=\frac{1}{2}\sqrt{\frac{\pi}{\theta}}\ \mathrm{erf}\left(\frac{\sqrt{\theta}}{\hbar}E\right).
\end{equation}
Going through the above calculation again, the transition probability becomes
\begin{equation}P(\nu_\alpha\rightarrow\nu_\beta)=\sin^2(2\theta)\sin^2\left(\frac{\Delta m^2}{4\hbar E}\exp\left(-\frac{\theta}{\hbar^2} E^2\right)t\right),\label{eq:mod_prob}\end{equation}
where the crucial step is given explicitly by
\begin{eqnarray}
\nonumber\omega(E_1)-\omega(E_2)&=&\frac{1}{2}\sqrt{\frac{\pi}{\theta}}\left(\mathrm{erf}\left(\frac{\sqrt{\theta}}{\hbar}\cdot p\,\sqrt{1+\frac{m_1^2}{p^2}}\right)-\mathrm{erf}\left(\frac{\sqrt{\theta}}{\hbar}\cdot p\,\sqrt{1+\frac{m_2^2}{p^2}}\right)\right)\\
\nonumber &\approx& \frac{1}{2}\sqrt{\frac{\pi}{\theta}}\left(\mathrm{erf}\left(\frac{\sqrt{\theta}}{\hbar} E\right)+ \frac{m_1^2}{\hbar\left|p\right|}\sqrt{\frac{\theta}{\pi}}\ e^{-\theta E^2/\hbar^2}\right)\\
\nonumber &\quad& -\frac{1}{2}\sqrt{\frac{\pi}{\theta}}\left(\mathrm{erf}\left(\frac{\sqrt{\theta}}{\hbar} E\right)+\frac{m_2^2}{\hbar\left|p\right|} \sqrt{\frac{\theta}{\pi}}\ e^{-\theta E^2/\hbar^2}\right)\\
&=&\frac{\Delta m^2}{2\hbar E}e^{-\theta E^2/\hbar^2}.\label{eq:approx}
\end{eqnarray}<
The extension of this model to the realistic three flavour case is slightly more complicated, but the general idea remains the same.
In \cite{Sprenger2011} we applied our model to data from the MINOS experiment.
The results are shown in figure \ref{fig:minos}.
As you can see, there is a modification of the oscillation pattern which, however, is very small for relevant values of $\sqrt{\theta}$.
However, in \cite{Sprenger2011}, we found scenarios in which the effect is more pronounced and could lead to a strong bound on the minimal length.
In particular, gamma-ray bursts and active galactic nuclei are candidates for the emission of ultra-high energetic cosmogenic neutrinos.
These neutrinos might have energies larger than $\hbar/\sqrt{\theta}$.
According to Eq.(\ref{eq:mod_prob}) these neutrinos would not oscillate, leaving a clear experimental signal.
\section{Results}
Both the GUP and the NCG approach presented in this article have been applied to a wide variety of problems from quantum mechanics and general relativity.
In this section, we want to give a small overview over what has been investigated in these frameworks.

Besides basic problems in quantum mechanics such as the harmonic oscillator \cite{Kempf1995} and the hydrogen atom \cite{Brau1999}, GUP has been extended to quantum field theories \cite{Kempf1997}, \cite{Hossenfelder2003} and used to derive modifications for basic scattering processes such as $e^+e^-$-annihilation where it turns out that the standard model cross section is reduced considerably \cite{Hossenfelder2003}.
Moreover, minimal length thermodynamics has been studied from which a modification for stable neutron star configurations was found in \cite{Wang2010}.

The coherent state approach to NCG has been applied mostly to black hole physics.
Non-commutative solutions have been obtained for all classical black hole solutions and show that the singularity that plagues the classical solutions is tamed.
Besides being regular everywhere, these solutions have a finite temperature throughout their evaporation, thereby getting rid of another divergence of the (semi-)classical theory \cite{Nicolini2008}, \cite{Modesto2010}, \cite{Banerjee2008}.
Moreover, black holes have been investigated in higher-dimensional space-time in relation to their phenomenology in collider experiments \cite{Rizzo2006}, \cite{Gingrich2010}, \cite{Nicolini2011}. 
Furthermore, the model was used to study the dimensionality of space-time at the fundamental scale which turned out to show a fractal behaviour \cite{Nicolini2010}, \cite{Modesto2009}.

Of course, this is just a small part of the literature, but this short list will provide a good overview for the interested reader.

\section{Conclusions}
In this article we have reviewed the GUP model and the coherent state approach to NCG as effective models for quantum gravity.
This allows us to make model-independent predictions as the appearance of a minimal length is supported by most current candidate theories.
These approaches include the effect of a minimal length and, due to their simplicity, offer a rich phenomenology.
Even though the effects are suppressed, the LHC and other near-future experiments will be able to put limits on the minimal length and, thereby, shed light on the nature of space-time.

\end{document}